\documentclass[10pt, conference, compsocconf]{IEEEtran}
\ifCLASSINFOpdf
\else
\fi

\usepackage{amsmath}
\usepackage{psfrag,graphicx}
\usepackage{amssymb}
\usepackage{algorithm}
\usepackage[noend]{algorithmic}
\usepackage{url}
\usepackage{subfigure}

\newtheorem{definition}{Definition}
\usepackage{balance}  

\begin{document}
%
\title{Scalable Cost-Aware Multi-Way Influence Maximization}


\author{\IEEEauthorblockN{Hong-Han Shuai}
\IEEEauthorblockA{National Taiwan University\\
d99942020@ntu.edu.tw}
\and
\IEEEauthorblockN{Hong-Han Shuai}
\IEEEauthorblockA{National Taiwan University\\
d99942020@ntu.edu.tw}
\and
\IEEEauthorblockN{Hong-Han Shuai}
\IEEEauthorblockA{National Taiwan University\\
d99942020@ntu.edu.tw}
\and
\IEEEauthorblockN{Hong-Han Shuai}
\IEEEauthorblockA{National Taiwan University\\
d99942020@ntu.edu.tw}
}

\maketitle

\begin{abstract}

\end{abstract}


\IEEEpeerreviewmaketitle

\section{Introduction}
Viral marketing is different from other marketing strategies since it leverages the influence power in intimate relationship, e.g., close friends, family members, couples. Due to the development and popularity of social networking services, such as Facebook, Twitter, and Pinterest, the new notion of ``social media marketing'' has appeared in recent
years and presents new opportunities for enabling large-scale and prevalent viral marketing online. To boost the growth of their sales, business is embracing social media in a big way. According to USA Today, the sales of software to run corporate social networks will grow 61\% a year and be a $6.4$ billion business by 2016\footnote{\url{http://usatoday30.usatoday.com/money/economy/story/2012-05-14/social-media-economy-companies/55029088/1 
}}. On the other hand, general advertisement channels such as TV, and newspaper are not dead yet. While there has been a significant drop since the rise of the Internet age, 116.3 million Americans have a television according to a Nielsen 2014 report.\footnote{\url{http://www.nielsen.com/us/en/insights/news/2014/nielsen-estimates-116-3-million-tv-homes-in-the-us.html}} Despite of the prevalence of social media marketing, TV is still an important traditional marketing method companies should consider advertising on. Advertising efforts on TV or radio has the benefit of reaching a mass audience. A recent line of research also focuses on how to model and quantify the influence from external (e.g., TV, Online News) and internal (i.e., friends, followees) exposures together \cite{MyersKDD12,EftekharKDD13}.

Consider the following scenario as a motivating example. A telecom firm attempts to market the new plan through multiple ways, e.g., TV commercial, social media, cold calls. The company has limited budgets on each advertising way such that it can only broadcast the TV commercials several times or select some initial users in the online network to adopt it (by giving them discount or free phone accessories). The ideal case of the company is that the TV audiences love the advertisement and adopt the new plan or selected seed users love the new plan. Afterward, the initial users start canvassing their friends for the new plan on the social network, and their friends will influence their friends' friends and so on, and thus through the viral marketing a large population in the social network would adopt the new plan. 

The above marketing strategy can be regarded as a combination of traditional media marketing and social media marketing. The problems are how many the budgets should be allocated to each advertising way and whom to select as the initial users so that they eventually influence the largest number of people in the network. Moreover, since the goal is to maximize the revenue, it is desirable to construct a precise cost model. Take the telecom case as an example, when users became the member of the company, the fees may be reduced due to the discounts on all calls made within the intra-network.

With this objective in mind, we formulate a new fundamental optimization problem, named \textit{Cost-Aware Multi-Way Influence maXimization (CAMAIX)}. The problem is given a social graph $G$, where each node represents a candidate person and is associated with an activation probability vector of traditional media, and each edge has a social influence probability to indicate the influence power between the two persons. Given the user-specified budget upper bound for each advertising way and the precise cost model, the goal of CAMAIX is to automatically allocate budgets and select seed users which maximizes the total revenue. 

There are three major challenges of CAMAIX: i) The spread maximization problem in the Independent Cascade (IC) model \cite{DavidKempe:KDD03} suffers from the expensive computation problem since the difficulty of the influence spread given a seed set is $\#P$-hard. Also, instead of finding an exact algorithm, Monte-Carlo simulations of the influence cascade model run a large number of times in order to obtain a correct estimation of the spread. ii) The number of seed nodes is non-fixed, which is different from traditional influence maximization problem and complicates the problem. Let $n$ denote the number of nodes in $G$. The enumeration approach for selecting $k$ seed nodes needs to evaluate $C^n_k$ candidate groups, whereas the enumeration approach for selecting non-fixed seed nodes is $2^n$.\footnote{It is worth noting that the cost of activating seed nodes is considered so that the total revenue will be reduced when wasting budgets on users who are really not willing to use.} iii) The budget allocation problem needs to deal with the interplay between different advertising ways and be extended to adopt more complicated real settings.

Aiming to efficiently solve the multi-way influence maximization with more sophisticated real settings, we systematically explore various heuristics, including \emph{Social Influence Pruning} (SIP) and \emph{Adaptive Budget Allocation} (ABA), to design our algorithm \emph{Intermediate Seeds Selection with Budget Allocation} (ISSBA). The idea of SIP is to incrementally construct the best seed set by maintaining a number of intermediate subsets. Therefore, by iteratively expanding the best intermediate seed sets from the subsets obtained in previous iterations, SIP finds seed sets with good quality efficiently. Also, we prove the performance bound of the proposed algorithm is $1/2$. On the other hand, given the upper bound of the budget for each advertising way, ABA efficiently calculates the optimal budget for each advertising way via dynamic programming.

The contribution of this paper is listed as follows.
\begin{itemize}
\item We formulate a new optimization problem, namely CAMAIX, to consider the
multi-way influence maximization with a sophisticated cost model, which is $\#P$-hard. To the best of the our knowledge, there is no real system or existing work in the
literature that efficiently addresses the issue of multi-way influence maximization based on
real settings.

\item We design Algorithm \emph{ISSBA} to find the solution to CAMAIX with an approximation ratio. Experimental results demonstrate that the solution returned by \emph{ISSBA} it perform the baseline algorithms in both solution quality and execution time on the large-scale datasets.
\end{itemize}

The rest of this paper is organized as follows. Section \ref{Prilim} formulates the CAMAIX problem and surveys related work. Section \ref{sec:03ISSBA} presents \emph{ISSBA} with social influence pruning and adaptive budget allocation. We report the experimental results in Section \ref{sec4:exp} and conclude this paper in Section \ref{conclu}.

\section{Preliminary\label{Prilim}}

\subsection{Problem Definition\label{IMProb}}
Let $S$ denote the seed set. A user $v$ has probability $p_{v,a}$ to be activated as a seed, i.e., $v \in S$, if the advertisement is sent to $v$ through a multiple advertising ways as follows.
\begin{equation}
1- \prod_{a \in A}(1-p_{v,a}).
\end{equation}
The activation probability $p_{v,a}$ of broadcasting advertisements $a$, such as TV and billboard, is the product of the probability that the advertisement $a$ broadcasts to user $v$ and the probability of user $v$ being activated by broadcasting advertisement. If a user $v$ is activated, the propagation starts from $v$ to his neighbors $u \in N_{u,v}$ with probability $p_{u,v}$. Moreover, let $c_{v,a}$ denote the cost $c_{v,a}$ for each advertising way $a \in \mathcal{A}$ on node $v$, where $\mathcal{A}$ is the advertise way set. Notice that $c_{v,a}$ for broadcasting advertisements $a$ is the total advertisement expense divided by the number of seed nodes $|S|$. 

Given a directed social network $G=(V,E)$, where node $v \in V$ and each edge $e_{i,j}\in E$ are associated with an initial fee $f(v)$ and an influence probability $p_{u,v}$ that user $u$ activates $v$ respectively, an advertising way set $\mathcal{A}$, and a budget $B_a$ and a cost $c_{v,a}$ for each advertising way $a \in \mathcal{A}$ on node $v$, this paper studies a new optimization problem called Cost-Aware Multi-Way Influence Maximization (CAMWIM) for finding the optimal budget and the seed set $S_p$ of vertices to maximize the revenue $R(S_p)$, i.e., $R(S_p)$

\begin{eqnarray*}
&=& \sum_{v \in V} ap(v,S_p,G)[f(v)-\sum_{u \in N_v}{d(v,u)ap(u,S_p,G)}]\\
&&-\sum_{a \in A}{\sum_{v \in S}c_{v,a}},\\
\end{eqnarray*}%
where $ap(v,S_p,G)$ and $d(v,u)$ represent the probability that the user $v$ will be activated with the seed set $S_p$ and the discount of activating $v$ related to its neighbor $u$, respectively. The discount is suitable for many different scenarios, such as telecom (intra-network free) and direct sale (agent commission), and is set as $0$ for no-discount cases.

\subsection{Related Work}
Influence maximization is to find a set of influential nodes, which are targeted as initial active nodes, to maximize the spread.   
The problem has been connected to the Independent Cascading (IC) model and the Linear Threshold (LT) Model models in \cite{DavidKempe:KDD03}. D. Kempe et al. \cite{DavidKempe:KDD03} show that the influence maximization problem is NP-Hard and propose a greedy algorithm for both IC and LT models, with the guarantee of the solution quality. However, the greedy algorithm needs Monte Carlo simulations to estimate the expected spread, which is time consuming. J. Leskovec et al. \cite{Leskovec:KDD07} proposes CELF to further speed up Monte Carlo simulations. Nevertheless, for large scale social networks, CELF is still not efficient enough. Several heuristic methods are proposed, such as degree discount \cite{WeiChen:KDD09}, PMIA \cite{WeiChen:KDD10}, and IRIE \cite{WeiChen:ICDM12}, to find initial active nodes very efficiently.

\section{Cost-Aware Influence Maximization\label{sec:03ISSBA}}
To tackle CAMAIX, a basic approach is to enumerate all possible seeds and combinations of budgets, and retrieve the one with largest revenue. However, the enumerative approach is not scalable since there are $2^n$ combinations for seed selection. To address the challenges, we propose a framework called Intermediate Seeds Selection with Budget Allocation (ISSBA) including \emph{Social Influence Pruning} (SIP) and \emph{Adaptive Budget Allocation} (ABA). SIP iteratively expands the best intermediate seed sets from the subset obtained in previous iterations. Moreover, we leverage the merit of MIA model to efficiently approximate the computation of Monte-Carlo simulation. Finally, ABA exploits dynamic-programming for allocating the budgets adaptively.

\subsection{Social Influence Pruning with Quality Guarantee}
Here, we describe our proposed SIP in detail. SIP first constructs all one-item subsets $\mathbb{S}_1^1, \mathbb{S}_1^2, ..., \mathbb{S}_1^{|V|}$, where each $\mathbb{P}_1^i$ contains exactly one item $i \in I$. The corresponding join group $F_1 ^i$ for each $\mathbb{P}_1^i$ are computed and the best $\kappa$ sub-packs with the largest join groups are reserved \footnote{The calculation process to obtain each $F_1^i$ is described in \ref{wp} later. For ease of understanding, we rename the best $\kappa$ sub-packs as $\mathbb{P}_1^1$, $\mathbb{P}_1^2$, ...$\mathbb{P}_1^\kappa$.}. IPO then generates 2-item sub-packs $\mathbb{P}_2^{i,j}$ by adding every possible items $j \in I - \mathbb{P}^i_1$ into the best $\kappa$ $\mathbb{P}^i_1$ separately. For example, IPO expands $\mathbb{P}_1^1$ into sub-packs $\mathbb{P}_2^{1, 1}, \mathbb{P}_2^{1, 2}, ..., \mathbb{P}_1^{1, |I|-1}$ by adding each item $j \in I - \mathbb{P}^i_1$ into $\mathbb{P}_1^1$. Note that during the generation of sub-packs, multiple sub-packs that contain the same items may be generated. IPO discards those additional duplicate sub-packs. Similarly, IPO computes the join group $F_2 ^{i,j}$ for each 2-item sub-packs $\mathbb{P}_2^{i,j}$ and reserves the $\kappa$ ones with the largest join groups for generating 3-item sub-packs \footnote{For ease of understanding, the best $\kappa$ sub-packs are renamed as $\mathbb{P}_2^1$, $\mathbb{P}_2^2$, ...$\mathbb{P}_2^\kappa$.}. The process runs iteratively until the $\sigma$-item sub-packs are generated and the best $\sigma$-item sub-pack is returned. The pseudo code of IPO is showed as Algorithm \ref{alg:poa}.

\subsection{Approximate Influence Maximization}
The spread maximization problem in the Independent Cascade
(IC) model \cite{DavidKempe:KDD03} suffers from the expensive computation problem since the difficulty of the influence spread given a seed set is $\#P$-hard. To efficiently address this issue, an approximate IC model,
called MIA, has been proposed \cite{WeiChen:KDD10,ChenAAAI2012}. The social influence from a person $u$ to another person $v$ is effectively approximated by their maximum influence path (MIP), where the social influence $w_{u,v}$ on the path ($u$,$v$) is the maximum
weight among all the possible paths from $u$ to $v$. MIA creates
a maximum influence in-arborescence, i.e., a directed
tree, MIIA($t$,$\theta$) including the union of every MIP to $t$ with
the probability of social influence at least $\theta$ from a set $S$ of leaf nodes. The MIA model has been widely adopted to describe the phenomenon of social influence in the literature with the following definition on activation probability, which is basically the same as the acceptance probability if $s$ broadcasts friending invitations to all nodes in $MIIA(t,\theta)$.

\begin{definition}
The activation probability of a node v in $MIIA(t,\theta)$ is $ap'(v, S,MIIA(t,\theta))$=

\begin{equation*}
\left\{ 
\begin{aligned}
&~~~~~~~~~~~~~~~~~~~~~~~~1\text{, if }v \in S~~~~~~~~~~~~~~~~~~~\\ 
&~~~~~~~~~~~~~~~~~~~~~0\text{, if }N^{in}(v)= \emptyset ~~~~~~~~~~~~~~~~\\
&1-\prod_{u \in N^{in}(v)}(1-ap'(v, S,MIIA(t,\theta))p_{u,v})\text{, otherwise,} \\
\end{aligned}\right.
\end{equation*}%
\end{definition}
Note that $ap'(u,S,MIIA(t,\theta))p_{u,v}$ is the joint probability that $u$ is activated and successfully influences $v$, and $u$ can never influence $v$ if it is not activated. Therefore, the activation probability of a node $v$ can be derived according to the activation probability of all its in-neighbors, i.e., the child nodes in the tree. Since $S$ is the set the leaf nodes, the activation probabilities of all nodes in $MIIA(t,\theta)$ can be efficiently derived in a bottom-up manner from $S$ toward $t$.
\subsection{Computation Reduction of Budget Allocation via Dynamic Programming}

\section{Experiment\label{sec4:exp}}
\subsection{Experiment Setting}
call detail records collected by a telecom operator
 
As [3] [4],  we study telecommunications social networks extracted from a large amount of Call Detail Records (CDRs).

\section{Conclusion\label{conclu}}
The conclusion goes here. this is more of the conclusion



The authors would like to thank...
more thanks here



%



\bibliographystyle{abbrv}
\bibliography{ref}

\begin{thebibliography}{1}

\bibitem{ChenAAAI2012}
W.~Chen, W.~Lu, and N.~Zhang.
\newblock Time-critical influence maximization in social networks with
  time-delayed diffusion process.
\newblock {\em AAAI}, 2012.

\bibitem{WeiChen:KDD10}
W.~Chen, C.~Wang, and Y.~Wang.
\newblock Scalable influence maximization for prevalent viral marketing in
  large-scale social networks.
\newblock In {\em KDD}, pages 1029--1038, 2010.

\bibitem{WeiChen:KDD09}
W.~Chen, Y.~Wang, and S.~Yang.
\newblock Efficient influence maximization in social networks.
\newblock In {\em KDD}, pages 199--208, 2009.

\bibitem{EftekharKDD13}
M.~Eftekhar, Y.~Ganjali, and N.~Koudas.
\newblock Information cascade at group scale.
\newblock In {\em KDD}, 2013.

\bibitem{WeiChen:ICDM12}
K.~Jung, W.~Heo, and W.~Chen.
\newblock Irie: Scalable and robust influence maximization in social networks.
\newblock In {\em ICDM}, pages 918 --923, 2012.

\bibitem{DavidKempe:KDD03}
D.~Kempe, J.~Kleinberg, and E.~Tardos.
\newblock Maximizing the spread of influence through a social network.
\newblock In {\em KDD}, pages 137--146, 2003.

\bibitem{Leskovec:KDD07}
J.~Leskovec, A.~Krause, C.~Guestrin, C.~Faloutsos, J.~VanBriesen, and
  N.~Glance.
\newblock Cost-effective outbreak detection in networks.
\newblock In {\em KDD}, pages 420--429, 2007.

\bibitem{MyersKDD12}
S.~Myers, C.~Zhu, and J.~Leskovec.
\newblock Information diffusion and external influence in networks.
\newblock In {\em KDD}, 2012.

\end{thebibliography}

\end{document}